\documentclass{article}

\usepackage{graphicx}

\begin{document}

\title{WWW Spiders: an introduction}

\author
{Massimiliano Zanin \\ massimiliano.zanin@hotmail.com}

\date{\today}

\maketitle

\begin{abstract}
In recent years, the study of complex networks has received a lot of attention. Real systems, including information networks and relationships between persons and users, have gained importance in scientific publications, despite of an important drawback: the difficulty of retrieving and manage such great quantity of information.

This paper wants to be an introduction to the construction of {\it spiders} and {\it scrapers}: specifically, how to program and deploy safely these kind of software applications. The aim is to show how software can be prepared to automatically surf the net and retrieve information for the user with high efficiency and safety.
\end{abstract}


\section{Introduction}

\subsection{Internet agents}

\begin{quotation}
{\em ``Agents are here to stay, not least because of their diversity, their wide range of applicability and the broad spectrum of companies investing in them. As we move further and further into the information age, any information-based organization that does not invest in agent technology may be committing commercial hara-kiri." }

From: Nwana, H.S. {\it Software Agents: An Overview.} Intelligent Systems Research AAandT, BT Laboratories, Ipswich, United Kingdom, 1996
\end{quotation}

Out there, there are about three billion web pages connected together, creating what we now know as the {\it Internet} or {\it World Wide Web}. Of course, this means a huge quantity of information that can be useful somehow, but it also means a problem: a human user cannot easily control and search in this too vast collection.

So, in this context, web agents were born, as personal software assistants with authority delegated from they users, that surfs the WWW with some task. 

There are many kind of web agents and a possible classification is shown below:

\begin{itemize}
    \item {\em Web robots, spiders and wanderers}: these are programs that cruise in the information space of Internet, searching for new information and resources, to create indices in the Webspace, and so on.
    \item {\em Web commerce agents}: a group of automated shoppers and comparison robots that help companies in smart on-line buying, trading or broking.
    \item {\em Worms and viruses}: malicious agents that replicate themselves in the background, traveling from machine to machine, via hardware (like floppy disks) or software connections (Internet).
    \item {\em MUD agents and chatterbots}: agents focusing on entertainment. They can interact with an user, answer inquiries, or simply chat with an human player.
\end{itemize}

In this paper, we will focus in how to construct and deploy the first kind of agents, because of their great interest and usefulness in scientific research: especially, in real complex networks study.

\subsection{Spidering and Scraping}

What is the difference between {\it spiders} and {\it scrapers}? Roughly speaking, both are programs, controlled by a human user, that surf the net to collect information. Moreover, we normally want to get this information in a different structured presentation, for further studies.

If we want to be more precise, {\it spiders} are programs that grab an entire web page, file, or collections of both; for example, the profile of each user in a chat community. On the other side, a {\it scraper} only grab a well-limited and specific part in the entire page: in the previous example, we may only want to know the age of each user.

As we will see, if it is possible, {\it scrapers} are better than {\it spiders}, since they need to download only little information, so they are generally faster, and they do not waste connection bandwidth.

\section{Agents ethic (and legality)}

When creating an agent to surf the web, or a given web-page, we must observe some rules, ethical and legal, in order to respect other users' rights: at the end, we are taking advantage of the work and the information of other people.

\subsection{Legal issues}

Often the legal aspect of an action in Internet is not clear: and agents are not an exception. There are many points related to web laws that are not well understood, and many aspects can change within different nations. The best practice to avoid legal problems is to ask and get a permission from the web master(s), especially with small (that is, pages run by a single person) pages.

Next step, is to have a look at the {\it Acceptable Use Policy} (AUP), the {\it Terms of Service} (TOS) or the {\it Terms of Use} (TOU) of the web site. Usually, you can find a link in the bottom part of the page, side to side with the copyright information; Yahoo! (see \verbºwww.yahoo.comº) has a link in the last part of their frontal page, while Google has it in the end of their About page (\verbºwww.google.com/intl/en/about.htmlº).

Just an example; the eBay web page makes clear what are the limitations that you must obey when using a robot in their site (at \\ {\verbºpages.ebay.com/help/policies/uapp.htmlº):

\begin{quotation}
{\em ``Access and Interference

The Site contains robot exclusion headers. Much of the information on the Site is updated on a real-time basis and is proprietary or is licensed to eBay by our users or third parties. You agree that you will not use any robot, spider, scraper or other automated means to access the Site for any purpose without our express written permission."} 
\end{quotation}

Of course, obey the Terms of Service of the web page is now enough; take intellectual property from a server to put it in another web is a violation of the copyright laws, and it should be avoided at any cost.

\subsubsection{Consequences}

What would be the consequence of creating a spider that goes against the TOS or that creates some malfunction in the target server?

The first response from a webmaster would be to block your IP; in the past, Google Labs blocked groups of IP addresses, because of an automated agent was accessing to their data violating the TOS.

Next step, may be a ``cease and desist" letter: if you don't desist from targeting the server, it may end in a lawsuit. Of course, this last possibility is very rare, at least if you don't attack the site with the intention of blocking its activities (what is called {\it denial of service}): nevertheless, the better solution is always to stop before this step.

\subsection{Robots.txt}

In years 1993 and 1994, a new problem started to grow up: robots were born, starting their activity of visiting web pages where they were not welcome. There are many reason to exclude a robot from a web site (especially if it has been badly programmed); certain agents send rapid-fire requests to the server, abusing of its bandwidth and slowing the access to other human users; on the other side, some robots may be travelling in parts of the Web where they should not stay, or getting sensitive information.

To solve this situation, in 1994 the {\it Standard for Robot Exclusion} was created \cite{Robs}. A webmaster that wants to control robots access, must create a {\it robot exclusion file} in the standard path \verbº/robots.txtº; when a robot visits a web-page, it should check for this file, and see what part it can visit. 

There are two main records in the {\it robots.txt} file:

\begin{itemize}
    \item {\it User-Agent}: the value of this field identify the robot (or a \verbº*º to apply the rule to any robot); if the robot recognize its name in this field, it must respect the prohibitions specified in the next section.

    \item {\it Disallow / Allow}: here the robot finds a partial string describing the prefix portion of the URL that should not visit (or that it can visit, for the {\it Allow} record).
\end{itemize}

The following is part of the {\it Robots.txt} file of the arXiv page (at \\ \verbºwww.arxiv.org/robots.txtº):

\begin{quote}
\begin{verbatim}
User-agent: *
Disallow: /cgi-bin/
Disallow: /e-print/
Disallow: /src/
...
User-agent: Googlebot
Allow: /archive
Allow: /year
...
Disallow: /cgi-bin/
Disallow: /e-print/
\end{verbatim}
\end{quote}

In this case, the webmaster has declared that any agent (the asterisk) should not enter a group of pages: the {\it Googlebot} agent is an exception, because it can access some directories (clearly to allow an indexing of the papers hosted).

When programming an agent, it is a good practice to respect the limitations of the {\it robots.txt} file, either dodging that site, or by implementing this limitation in the robot itself (a sample Perl code can be found at: \verbºwww.cpan.org/º).

\subsection{The Seven Robots Commandments}

We have seen in the previous sections how there are some rules (or laws) that should be respected when deploying a robot. There are other {\it suggestions} that you should consider \cite{Comman}: 

\begin{enumerate}
    \item {\em Thou Shalt Announce thy Robot}: when deploying a robot, it is a good practice to announce it to the target server administrator. This will help him to minimize the impact on the performance of the site, and moreover he/she may tell you how to get some information. A part from that, it is useful to contact with the administrator before creating a robot: maybe he/she can give you the information you are looking for without further work.
    \item {\em Thou Shalt Test, Test and Test thy Robot Locally}: the perfect solution to test your program, is to create some web-servers locally, and test the creation in it; if this is impossible, try to check every possibility of the program, to ensure a perfect performance.
    \item {\em Thou Shalt Download in Moderation}: always ask and retrieve the minimum quantity of information needed; when studying the HTTP protocol, we will see tricks to minimize the size of the information passed.
    \item {\em Thou Shalt Keep thy Robot Under Control}: Never launch a robot and forget about it! Someone should stay close to check the robot progress, and detect and correct any problem that could happen; just an example: never let a robot enter a closed loop - it will download the same group of pages {\it Ad infinitum}.
    \item {\em Thou Shalt Stay in Contact with the World}: when the robot is out working for you, make sure that the webmaster (or other users) can contact you easily (in the client options, we will see how to specify an e-mail address of the creator of the robot).
    \item {\em Thou Shalt Respect the Wishes of Webmasters}: contact the webmaster of the site, and respect his/her instruction (what pages to visit, with which frequency, and so on); on the other side, trying to convert him to your cause may not be a good idea: probably, he/she won't be interested in your work.
    \item {\em Thou Shalt Share Results with thy Neighbors}: if you have worked hard to construct a robot to collect some kind of information around the Web, it is a good idea to share that data: other people can find it useful (and less webmasters will be bothered!).
\end{enumerate}

\section{Structure of the basic HTTP transaction}

The basic way to interact with a server, and therefore the main way to obtain information from a web page, is by using the HTTP protocol. Every time we surf a web page, a number of standard messages are exchanged between the client (i.e. the browser) and the server (the system managing the site). Those messages are sent using the TCP protocol \cite{TCP}, a communication protocol developed in 1974 for exchange data between two computers in a secure way.

\begin{figure}[!th]
\centerline{
\includegraphics[width=9cm]{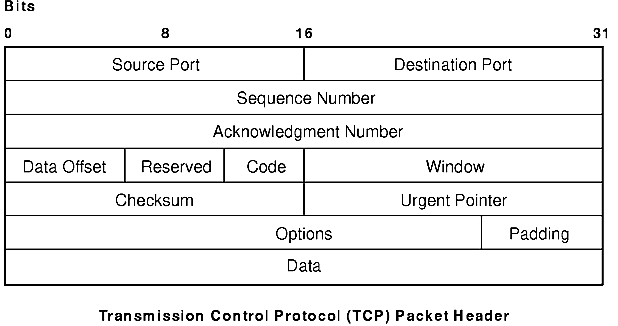}
}
\caption[sdh]{
Structure of a TCP packet} 
\label{fig:ImgPacket}
\end{figure}

The structure of a TCP packet (the smallest piece of information that can be sent to the receiver) is shown in Fig. \ref{fig:ImgPacket}. The good part, is that the Operating System handles this standard \footnote{In Windows {\texttrademark} environments, the library WinSock.dll is the responsible of managing any TCP communications.}, so the programmer doesn't need to know exactly how it works; nevertheless, a couple of concepts have to be clarified:

\begin{itemize}
    \item {\it IP address}: when sending an information to another user, in computer systems as in real life, we have to know the address of the destination; the {\it IP address} \cite{IPA} is just that: a group of four numbers defining an agent in Internet (i.e. $164.12.123.65$).
    \item {\it Port number}: each agent (or computer) connected via the TCP protocol has a number of {\it ports} where the message can be sent to; each message is sent from a port of the sender to a port of the receiver. The port number is a 16-bit number, so there are 65536 available ports; some of them have a fixed meaning, and should be used for a specific kind of communication: FTP on the port 21, SSH on 22, Telnet on 23, SMTP on 25 and HTTP on 80 \cite{Ports}.
    \item {\it Checksums}: when a message is sent, an error detection code is attached, so that the possibility of receive bad (or corrupted) information is very low.
\end{itemize}

On the TCP standard (that, at the end, is simply an {\it utility} for sending information) many other application level protocols are stacked, each one developed for a specific application: HTTP for web surfing (the most important when creating {\it spiders} or {\it scrapers}), FTP for file transfer, POP3 and SMTP for e-mail managing, IRC for on-line chatting, and so on (see Fig. \ref{fig:ImgTCP}).

\begin{figure}[!th]
\centerline{
\includegraphics[width=9cm]{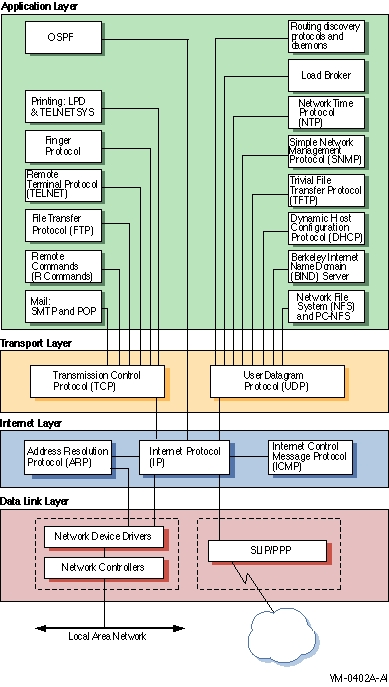}
}
\caption[sdh]{
Different layers in a network communication} 
\label{fig:ImgTCP}
\end{figure}

HTTP protocol \cite{HTTPP} has a very simple stateless structure (see Fig. \ref{fig:ImgCSCom}):

\begin{itemize}
    \item The client sends a {\it request} to the server, asking for some kind of information;

    \item Next, the server processes the request, and generates a {\it response} with the proper information.
\end{itemize}

This is the basic structure, and in most cases it's all that is needed to construct a Spider; nevertheless, some special circumstances can be found, like page redirection, authentication, cookies, and so on.

\begin{figure}[!th]
\centerline{
\includegraphics[width=7cm]{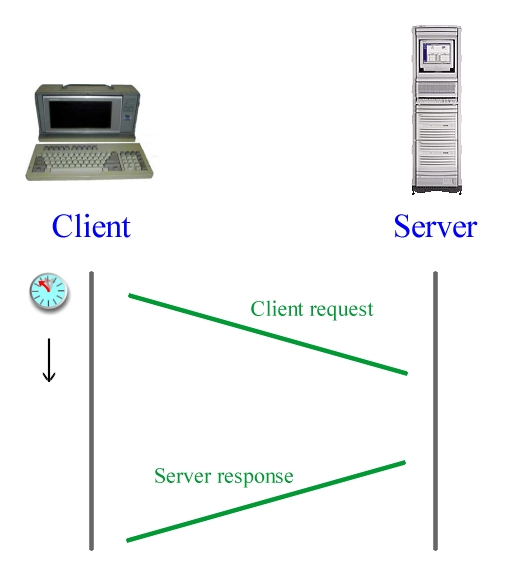}
}
\caption[sdh]{
Structure of a HTTP transaction} 
\label{fig:ImgCSCom}
\end{figure}

\subsection{Sending a simple request}

Following the scheme of Fig. \ref{fig:ImgCSCom}, the client must start the transaction, by sending a request to the server in the correct port (by default, the port for HTTP messages is 80). The basic request has the following structure:

\begin{itemize}
    \item First, we must specify the command for the server. The most common is the {\it GET} statement, that allows downloading a specified resource; later on, other commands will be analyzed, like {\it HEAD}, {\it POST} or {\it TRACE}.

    \item Next, we must tell the server what resource we want: that is, the URL of the item we want to download.

    \item The last part is which version of the HTTP standard we are able to manage. There are three options: {\it HTTP 0.9}, {\it HTTP 1.0} and {\it HTTP 1.1}; most of the servers and clients work with version 1.0: anyway, 0.9 is maintained for backward compatibility and 1.1 is starting to been used in many sites.
\end{itemize}

\subsection{Example 1: Basic image download}

To apply the basic request message of a client, we construct a program that simply wants to download an image: for example, the Google {\texttrademark} main image (at \verbºhttp://www.google.com/intl/en_ALL/images/logo.gifº).

This would be the request we need to send to the server:

\begin{quote}
\begin{verbatim}
GET /intl/en_ALL/images/logo.gif HTTP/1.0
\end{verbatim}
\end{quote}

The Google server IP is $72.14.207.99$ (to get the IP direction from a URL, check one of the many web pages that offers this utility), so that message should have been sent to port 80 of that IP. In response, the server sends us a message containing the image we wanted: so, the last action is to identify and save in a file the information needed and the task will be over.

This would have been the basic way to send the message to the target server. Nevertheless, this is not the best way to program a spider since in most cases, we have a compiler or a library formatting the message for us.
For example, this would be the source code in {\it Borland C++ Builder \texttrademark \cite{Borland}} to perform the same task:

\begin{quote}
\begin{verbatim}
MyHtml->Body = "Picture.gif";
MyHtml->Get( 
   "http://www.google.com/intl/en_ALL/images/logo.gif" 
   );
\end{verbatim}
\end{quote}
The first line tells the compiler that we want the response of the server in the file \verbº"Picture.gif"º, and the second retrieve the resource from the server.

\subsection{A server response}

In the previous section, we asked the server for an image. If we work with some HTML libraries that automatically parse the data, we would already saved that image; on the other side, if we are working directly with the message, we have to extract the information we are looking for.

The server message would have been something like this:

\begin{quote}
\begin{verbatim}
HTTP/1.0 200 OK
Content-Type: image/gif
Last-Modified: Wed, 07 Jun 2006 19:38:24 GMT
Expires: Sun, 17 Jan 2038 19:14:07 GMT
Server: gws
Content-Length: 8558
Date: Fri, 12 Oct 2007 18:10:56 GMT
Connection: Keep-Alive

GIF89a\x14\x01n\0÷\0\0÷÷÷ÿûÿçççÖÓÖïëïÎËέ\x14\0ÞÛÞ\x18E­\x18Iµ\x
104"\x10<"Æ\x18\0µ²µ÷ó÷Œ\x10\0Æ3/41/21/2º1/2\x18MÆçãçïïïÆÃÆ÷óï1/23/41/2ÆÇÆ\bQ\
bÎÏÎ\b$c!YÖÖ$\bÖ×Ö\x18Q̞œïº\0ÞßÞ\0e\0 ...
\end{verbatim}
\end{quote}

More generally, the server response has a structure like this:

\begin{quote}
HTTP {\it version Status-code Reason-phrase} \\
{\it Response header} \\
{\it - Empty line -} \\
{\it Response body}
\end{quote}

\begin{itemize}
    \item HTTP Version; as in the client request, we can expect any of the following options: {\it HTTP 0.9}, {\it HTTP 1.0} and {\it HTTP 1.1}. Normally, the server would try to answer with the version specified in the client request.

    \item Status code: a 3 digit numerical code that codifies the status of the operation; a $200$ answer is that the client request has been fulfilled. For a complete list of response codes, see Tabs \ref{tab:TabRC}-\ref{tab:TabRC5xx}.

    \item Reason phrase: a phrase that decodes the meaning of the Status code.

    \item Response header: in this optional part, the server can add a quantity of extra information useful to correctly interact with it. In the next section, we will see in details the structure of this item.

    \item Response body: here the resource asked by the client is attached; in the previous example, the server adds the content of the GIF file the client asked for.
\end{itemize}

\section{More HTTP options}

In section 3 we have seen the general structure of a HTTP transaction: a basic request from a client, and the response of the server. In both messages, the sender can add a {\it header}, that includes a list of information useful to interact with the other agent of the communication.

In this section, we are going to analyze the most important statements in both {\it request} (send by the client) and {\it response} (send by the server) {\it headers}, and we will see how they can help us in optimizing and improving the interaction with the web site.

\subsection{The HEAD command and up-to-date information}

Suppose that we find an web-page that is connected with a webcam of some interesting place, and that we need to download a collection of shots of that place; for example, at \\ \verbºhttp://http://www.monroecounty.gov/airport-camviewer.phpº you can see the image of the {\it Greater Rochester International Airport}, updated each 5 min. aprox. From the source code of the page, we can see that the URL of the image is \\ \verbºhttp://www.monroecounty.gov/airport/airport_00329.jpgº.

\begin{figure}[!th]
\centerline{
\includegraphics[width=6cm]{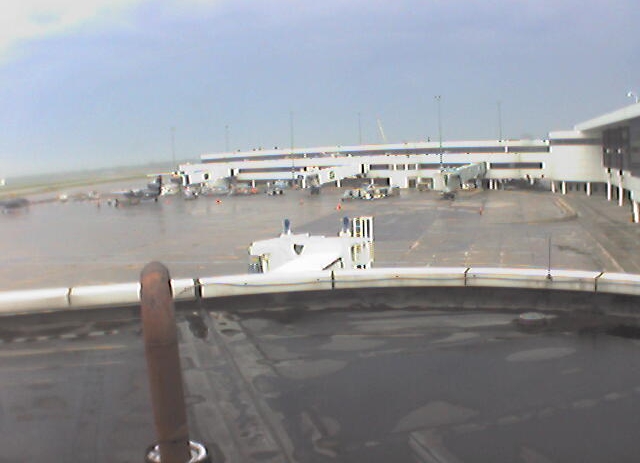}
}
\caption[sdh]{
A {\it Greater Rochester International Airport} webcam capture.} 
\label{fig:ImgAirport}
\end{figure}

Now, how could we download a series of images from that webcam? The easiest way, would be to make a program that downloads that image every 5 minutes. Nevertheless, it isn't very efficient; if the web page has some trouble updating the image, and the new one is available only after one hour, the program would have downloaded 20 identical images: that is, a great bandwidth would be lost without gaining information. Of course, the download of a JPEG image is not so costly: but in a more general case, this situation can happen with heavier resources.

A better solution is to make the transaction more intelligent: download the image only if it has changed. To make this, we need a protocol to ask the server if the resource has been updated, and then take actions according with the response.

\subsubsection{HEAD command}

We have already seen that the GET command allow us to download a specified resource from a server; now, we introduce a new command, the HEAD.

When sending to the server of the airport (\verbºwww.monroecounty.govº) the following message

\begin{quote}
\begin{verbatim}
HEAD /airport/airport_00329.jpg HTTP/1.0
\end{verbatim}
\end{quote}

we get the following response:

\begin{quote}
\begin{verbatim}
HTTP/1.1 200 OK
Date: Sat, 13 Oct 2007 02:23:08 GMT
Server: Apache/2.0.54 (Fedora)
Last-Modified: Thu, 10 May 2007 12:11:10 GMT
ETag: "1783ac-7d3b-8bd43380"
Accept-Ranges: bytes
Content-Length: 32059
Cache-Control: max-age=3600
Expires: Sat, 13 Oct 2007 03:23:08 GMT
Connection: close
Content-Type: image/jpeg
\end{verbatim}
\end{quote}

The server response is very similar to the one obtained with the GET command, but with an important difference: the server sends us information about the resource, but not the resource itself. This is very useful when we want to know what we are going to download prior to actually download the document, and therefore save bandwidth: for example, in our case we want to know if the image has changed before getting it.

\subsubsection{{\it Last-Modified} header}

As its own name stated, the {\it Last-Modified} header give us the date and time of the last modification of the resource. Although optional, the great part of the servers includes this information in their headers.

The way to manage this information is quite simple. First, get the header: if the {\it Last-Modified} time has changed with respect of the last request, download the resource; if not, wait a certain time.
This process can be seen in the following UML source code:

\begin{quote}
\begin{verbatim}
void download_timer (void)  
{  
   download HEAD
   if HEAD::LastMod distinct from LastMod 
   {
      download Resource
      LastMod = HEAD::LastMod
   }
}  
\end{verbatim}
\end{quote}

\subsubsection{ETags}

Sometimes, the {\it Last-Modified} header is not enough to retrieve information without waste of bits; for example, you may want to download several documents from a web page, all with the same name: as you don't know whether they are the same or not, you should download every copy of that document. These kinds of situations are resolved with the {\it ETag} (or {\it Entity tag}).

An {\it ETag} is a unique identifier that the server calculates and associates with every copy of a document; if the content is changed, then the {\it ETag} is changed too: therefore, this header can be efficiently used to identify resources that have changed.

The {\it ETag} header is defined in the HTTP 1.1 protocol, so not every server supports this method. If the server has been programmed to do so, it will automatically return a field in the Header section, like in the previous example:

\begin{quote}
\begin{verbatim}
HTTP/1.1 200 OK
...
ETag: "1783ac-7d3b-8bd43380"
...
\end{verbatim}
\end{quote}

\subsection{Other server headers}

In the following some of the most used server headers are listed; in addition these headers are also the most related with the construction of a spider. As a reference, the complete list of headers is reported in Tables \ref{tab:GHeader} - \ref{tab:EHeader} .

\subsubsection{Date}

As its name indicates, the Date header represents the date and time at which the message was originated by the server. The value passed is a HTTP date field, which looks like the following:

\begin{quote}
\begin{verbatim}
Date: Sat, 13 Oct 2007 02:23:08 GMT
\end{verbatim}
\end{quote}

\subsubsection{Retry-After}

Sometimes, the server can be not ready to answer to the client request, i.e. for maintenance of the infrastructure; in that cases, the response will be a {\it 503 Service Unavailable} message (see Table \ref{tab:TabRC5xx}). The {\it Retry-After} header indicates how long the service is expected to be unavailable; it can be a full HTTP date, or an integer number of seconds after the time of the response.

Here there are two examples:

\begin{quote}
\begin{verbatim}
Retry-After: Sat, 20 Oct 2007 10:00:00 GMT
Retry-After: 120
\end{verbatim}
\end{quote}

\subsubsection{Content-Length}

The {\it Content-Length} header field is used to tell the client the size, in bytes, of the resource asked to the server. If we are using the HEAD method, the field indicates the size of the item that would have been sent in response to a GET method.

In the previous example of the {\it Greater Rochester International Airport}, the JPG file had a total size of 32059 bytes.

\subsubsection{Content-Type}

The {\it Content-Type} header field tells the client what media type has been sent in answer to the request. Some examples follow:

\begin{quote}
\begin{verbatim}
Content-Type: text/html; charset=ISO-8859-4
Content-Type: audio/basic
Content-Type: video/mpeg
Content-Type: image/jpeg
\end{verbatim}
\end{quote}

\subsection{The problem of being identified}

As we have seen, when the server sends us the information we asked with the client, it also sends us other useful information, the {\it Response Headers}. The same occurs with the client: in a request, we can add a group of {\it headers} that carries data about us and about what we need from the web page.

{\it Client headers} have two main objectives:

\begin{enumerate}
    \item Identify who is the client (authentication);
    \item Allow the server to answer with customized information: as an example, if we are saying that our web explorer is Netscape, the server can prepare the HTML page to better fit in that program window.
\end{enumerate}

Following are some important {\it headers} and how they should be used.

\subsubsection{Who am I? the From header}

The {\it From} header is an optional field that should contain an Internet e-mail address of the person responsible of the client. For example, this would be a more complete request for the {\it Greater Rochester International Airport} page:

\begin{quote}
\begin{verbatim}
HEAD /airport/airport_00329.jpg HTTP/1.0
From: MyName@MyHost.com
\end{verbatim}
\end{quote}

When the administrator of the web page sees some problems about how a client is actuating, he/she can detect who is controlling that client and contact him. This is particularly useful when working with robots: the responsible of the web robot can be contacted if some problems occur on the receiving end.

\subsubsection{What I am using? the User-Agent header}

This {\it header} is similar to the {\it From header}: it permits identify who's generating the request, specifically with which program we are working.
This information is for statistical purposes of the web site, for tracing protocols violations, and for the automated generation of customized HTML codes.

Normally, the name of the Web Browser (or the name of the Spider) should be specified in this field. There are cases, however, when we may decide to skip this rule.
Let's make an example: the arXiv web page (at \verbºarxiv.orgº). This server stops any request coming from an unknown client; the following is an example of a request and its response:

\begin{quote}
\begin{verbatim}
GET /index.html HTTP/1.0
User-Agent: MyRobot
\end{verbatim}
\end{quote}

\begin{quote}
\begin{verbatim}
<!DOCTYPE HTML PUBLIC "-//W3C//DTD 
      HTML 4.01 Transitional//EN"
   "http://www.w3.org/TR/html4/loose.dtd">
<html>
<head><title>403 Forbidden</title></head>
<body>
<h1>Access Denied</h1>

 <p>Sadly, your client does not supply a proper User-Agent,
 and is consequently excluded.</p>
 <p>We have an inordinate number of problems with automated
 scripts which do not supply a User-Agent, and violate the
 automated access guidelines posted at arxiv.org
 -- hence we now exclude them all.</p>
 <p>(In rare cases, we have found that accesses through 
 proxy servers strip the User-Agent information. If this
 is the case, you need to contact the administrator of your 
 proxy server to get it fixed.)</p>


<p>If you believe this determination to be in error, see
<b>arxiv.org/denied.html</b> for additional 
 information.</p>
</body>
</html>
\end{verbatim}
\end{quote}

The problem is clear: this server doesn't want robots! Nevertheless, you can try (at your own responsibility) to overcome this problem, by telling the server that you are using a {\it normal} browser. For example, this request would generate a correct response:

\begin{quote}
\begin{verbatim}
GET /index.html HTTP/1.0
User-Agent: Internet Explorer
\end{verbatim}
\end{quote}

The following is a little list of classical agents:

\begin{quote}
\begin{verbatim}
Mozilla/4.0, Mozilla/5.0, Mozilla/6.0
Ace Explorer, AOL 8.0, curl/7.9.8
Explorer /5.02, FireFox
Internet Explorer 5x, Java1.4.0_02
\end{verbatim}
\end{quote}

A complete list of agents can be found on \\ {\verbºhttp://www.user-agents.org/º} or at \\ {\verbºhttp://www.pgts.com.au/download/data/browser_list.txtº}.

\subsection{Other client headers}

\subsubsection{If-Modified-Since}

When using the GET method, there are ways to optimize the quantity of information sent in the Web. We have already seen some strategies from the server side, now we will study some client side headers, like the {\it If-Modified-Since} header.

As its name states, we are telling the server to send us the resource only if that resource has changed since a certain date and time; normally, we will ask if there are modifications since the last time we have downloaded that item.
Following, there is an example of what will be the implementation in the case of the {\it Greater Rochester International Airport} page:

\begin{quote}
\begin{verbatim}
GET /airport/airport_00329.jpg HTTP/1.0
If-Modified-Since: Thu, 10 May 2007 12:11:10 GMT
\end{verbatim}
\end{quote}

The server can respond with two possibilities:

\begin{itemize}
    \item If the resource has changed, it will send us the image as usual;

    \item In the other case, the answer will be something like:

\begin{quote}
\begin{verbatim}
HTTP/1.0 304 Not Modified 
Last-Modified: Thu, 10 May 2007 12:11:10 GMT 
Content-Type: image/jpeg 
\end{verbatim}
\end{quote}

    The 304 server response indicates that the resource has not be changed since the date that we have specified, so we don't need the information right now.

\end{itemize}

\subsubsection{If-None-Match}

The {\it If-None-Match} header has the same function as the previous one, {\it If-Modified-Since}, but with an important difference: it works with ETags.
We can ask a resource to a server, and tell him to send us the information only if it has changed, by the following request:

\begin{quote}
\begin{verbatim}
GET /airport/airport_00329.jpg HTTP/1.0
If-None-Match: "1783ac-7d3b-8bd43380"
\end{verbatim}
\end{quote}

The server can send us two kind of responses: the new resource, or tell us that it has not changed (with a 304 response).

\subsubsection{Range}

We may face the situation where not the whole resource is needed: for example, in a HTML page, we may only want the first part of a very large HTML code (where the really interesting data is located). In this case, if the server support this option and the communication is via the HTTP 1.1 standard, we can use the {\it Range} header within the request, so that the server send us only the part of the document that we need.

Within any server response, if the following {\it Accept-Ranges} header is included, it signifies that the server accepts the {\it Range} option:

\begin{quote}
\begin{verbatim}
HTTP/1.0 200 OK 
...
Accept-Ranges: bytes 
\end{verbatim}
\end{quote}

Now, we only want the first 250 bytes of the document:

\begin{quote}
\begin{verbatim}
GET /afile.html HTTP/1.0
...
Range: 0-250
\end{verbatim}
\end{quote}

In the response, the server will tell us what part of the document is sending to us, and the total size of the resource itself:

\begin{quote}
\begin{verbatim}
HTTP/1.1 200 OK
...
Content-range: 0-250/152000
\end{verbatim}
\end{quote}

\section{Sending information}

Till now, we have only downloaded simply structures from the web, that needed no information to be generated: an image, maybe a page that automatically updates itself, and so on. In many cases, it would be enough: but, sometimes, we must interact further with the server.

As an example, we will take the {\it World airport codes} web page (at \\ \verbºwww.world-airport-codes.comº); we have a list of airport codes, and we want to get the distance between those airports all over the world.

\subsection{Codify the information in the URL}

To understand how a page like this works, we manually make an example: i.e. we ask the distance between LGA (La Guardia Airport, NYC) and CPT (Cape Town International, South Africa). When we get the answer, we will see that the address bar of our browser has changed to this: \\ {\verbºwww.world-airport-codes.com/dist/?a1=lga&a2=cptº (see Fig. \ref{fig:ImgMap}).

\begin{figure}[!th]
\centerline{
\includegraphics[width=7cm]{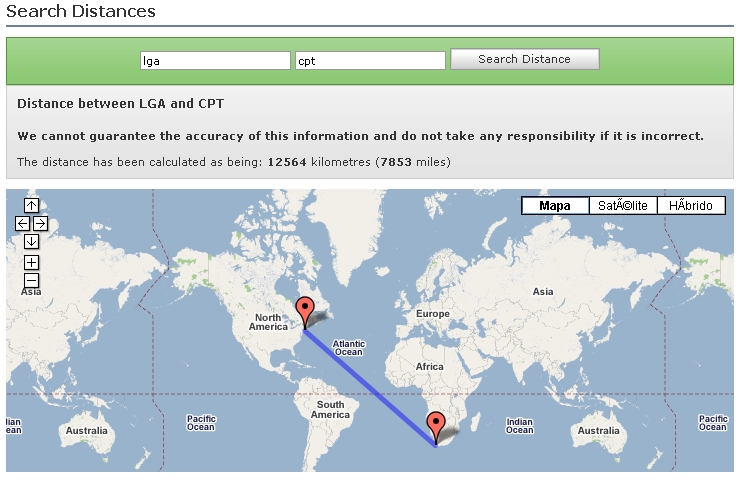}
}
\caption[sdh]{
Representation of the distance from LGA and CPT.} 
\label{fig:ImgMap}
\end{figure}

What the browser is doing, is codify the input information in the following way:

\begin{itemize}
    \item First, the normal URL of the page with the information (in our case, \verbºwww.world-airport-codes.com/dist/º);
    \item The \verbº?º symbol, telling that the following will be the parameters of the request;
    \item Every parameter needed: its name, the \verbº=º symbol, and its value (in our case, the parameter {\it a1} is the first airport, and its value is {\it LGA}, {\it La Guardia Airport};
    \item Each parameter is separated by a \verbº&º;
\end{itemize}

Although this is a very simple example, the 80\% of the pages you will find work in this way. If you want to make a program that collect distances between airports, you only need to GET the correct URL, by giving in the {\it a1} and {\it a2} parameters the codes of the airports. For example, you can directly get the distance between (\it LGA} and {\it BER} (the code of the Berlin airport, Germany), by GET the {\verbºwww.world-airport-codes.com/dist/?a1=lga&a2=berº} address.

\subsection{Obfuscated URLs}

Sometimes, the person who programmed the web site may have considered that showing the complete web address in the address bar is not good-looking. For example, you may open Wikipedia (at \verbºen.wikipedia.orgº) and search for a word, for example {\it home}: you will be redirect to the URL \\ \verbºhttp://en.wikipedia.org/wiki/Homeº, but there's no information on how the server has received the information about your search.

In these cases, you have two possibilities:

\begin{itemize}
    \item Extract the HTML code from the page, and try to understand which URL is called and with which parameters; although correct, this way is very long, and a good knowledge of HTML is needed.

    \item Use some external trick: at the end, the browser is sending the information, so there should be some way to intercept the data sent.
\end{itemize}

This second point can be achieved with an external program that analyzes the communications with the exterior world; for the Windows\texttrademark OS, there's a little utility called {\it TamperIE} (you can freely download it at \\ \verbºhttp://www.bayden.com/dl/TamperIESetup.exeº) that makes this job.

\begin{figure}[!th]
\centerline{
\includegraphics[width=9cm]{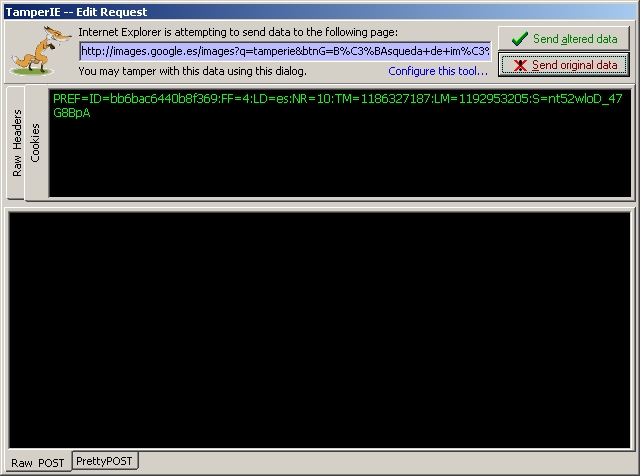}
}
\caption[sdh]{
A sample screen of {\it TamperIE}} 
\label{fig:ImgTamperIE}
\end{figure}

When making the search, a window of the program will appear, telling us that the browser is sending information to this web address: \\ \verbºhttp://en.wikipedia.org/wiki/Special:Search?search=homeº; in this way, you can easily see how a quest is made, so it can be implemented with no cost in a robot.

\subsection{The POST command}

Up to now, we have seen that the GET command has to be used when we want to retrieve information from a server, and that further parameters should be included in the URL address. Nevertheless, the HTTP standard specifies another command to get resources: the POST command.

{\it When should the POST or the GET commands be used?} There is no real rule about this point, but the official recommendation is to use the GET method when the request is {\it idempotent}: that is, when we make $N$ requests and the result is the same as for the first; or, in other words, when the request doesn't modify the internal status of the server. Moreover, we will see how the POST method is more suitable when many parameters are needed (the URL maximum length is 2083 characters).

In any case, from the client point of view, the previous question is not important, since it is the creator of the server that chooses the method he prefers. By using a tool like {\it TamperIE}, or by looking into the HTML code, a robot creator should see what command he/she must use, and apply it.

In the previous section we have seen how to get the distance between two airports with the GET method, codifying the information within the URL {\verbºwww.world-airport-codes.com/dist/?a1=lga&a2=cptº. What would be the TCP message if the POST method should be used?

\begin{quote}
\begin{verbatim}
POST /dist/ HTTP/1.0
Content-type: application/x-www-form-urlencoded
Content-length: 13

a1=lga&a2=cpt
\end{verbatim}
\end{quote}

There are two main differences:

\begin{enumerate}
    \item The parameters are not encoded in the URL: are attached at the end of the message, after a blank line, and without the \verbº?º symbol.
    \item Two headers should be added, telling the server how the information is passed (in this case, in the same code that the URL), and the total length of the information (in characters).
\end{enumerate}

\section{HTML Structure}

The great part of the information that we may want to collect in Internet is codified in HTML pages: so, in this section, we will have a look at the basis of the structure of the HTML documents, in order to understand the data in it (and eventually extract it automatically).
HTML is not dependent on the HTTP or the WWW: in fact, is a standard that exists by its own, and it is used to codify hypertextual e-mail, news, and so on. On the other side, a server can store information in any format, and then convert the data to a HTML structure on-the-fly.

A HTML document is based in a set of tags that codify the logical structure of the document, and information about how it should be displayed. Every element has a start tag (where the element name and properties are specified), the content, and the end tag. Each tag is enclosed by the \verbº<º and \verbº>º symbols, giving the following structure:

\begin{quote}
\begin{verbatim}
<Element>
   The content of the element.
<\Element>
\end{verbatim}
\end{quote}

\subsubsection{Head and body}

An HTML page normally has the following global structure:

\begin{quote}
\begin{verbatim}
<HTML>
   <HEAD>
      <TITLE>Sample page.<\TITLE>
   <\HEAD>

   <BODY>
      Some content...
   <\BODY>
<\HTML>
\end{verbatim}
\end{quote}

The HTML content is divided between a HEAD (which includes general information about the document: title, date of creation, and so on) and a BODY (where the specific information is codified).

Inside the body, the information is structured using the following tags.

\subsubsection{Headers}

To divide the document in sections, the following headers should be used:

\begin{quote}
\begin{verbatim}
<H1> This is the Header 1 <\H1>
<H2> This is the Header 2 <\H2>
<H3> This is the Header 3 <\H3>
\end{verbatim}
\end{quote}

\subsubsection{Paragraphs}

An example of a text paragraph:

\begin{quote}
\begin{verbatim}
<P>
   Here you should put the text
   for this paragraph.
\end{verbatim}
\end{quote}

\subsubsection{Lists}

An example of an unordered list:

\begin{quote}
\begin{verbatim}
<UL>
<LI> An element
<LI> An element
<\UL>
\end{verbatim}
\end{quote}

Note how the end tag is not needed in this case.

An example of an ordered list:

\begin{quote}
\begin{verbatim}
<OL>
<LI> First element
<LI> Second element
<\OL>
\end{verbatim}
\end{quote}

\subsubsection{Character highlighting}

\begin{quote}
\begin{verbatim}
<P> An example of character highlighting:

   <I> italics <\I>
   <B> bold <\B>
\end{verbatim}
\end{quote}

\subsubsection{Hyperlinks}

This is an example of how to include hypertextual links in the document. The page linked is \verbºwww.mypage.comº while the text that appears is {\it My Page}:

\begin{quote}
\begin{verbatim}
<A href="http://www.mypage.com"> My Page </A>
\end{verbatim}
\end{quote}

\section{Real examples}

In the previous sections, we have studied the background of an HTTP communication, how to contact a server, specify options, and so on. Nevertheless, normally you would not directly control the messages: it is more practical to use some libraries that automatize those processes, letting your attention to center on the other parts of the program.

In the following, we will see some example programs created in {\it C++} (specifically {\it Borland C++ Builder} \texttrademark), {\it Perl} and {\it Java} languages, using libraries to manage the HTTP standard (respectively, the {\it TNMHTTP} component and the {\it libwww-perl} package \cite{KHem}). Many other packages can be found in the Internet: apart from small differences, the underlying concepts are the same for everyone.

\subsection{Direct C++ TCP communication}

Following the example of the {\it Greater Rochester International Airport}, we want to write a program in C++ to download the HEAD part of the image using directly the TCP layer. This is not an HTTP library example, but it can be useful to see how a TCP communication is made.

\begin{quote}
\begin{verbatim}
char buffer[10000];

   ClientSocket1->Host = "www.monroecounty.gov";
   ClientSocket1->Port = 80;
   ClientSocket1->Open();

  TWinSocketStream *pStream = 
      new TWinSocketStream(ClientSocket1->Socket, 60000);

  pStream->WaitForData(1000);
  strcpy(buffer, 
     "HEAD /airport/airport_00329.jpg HTTP/1.0\n\n");
  pStream->Write(buffer, strlen(buffer) + 1);

  pStream->WaitForData(10000);
  pStream->Read(buffer, 9999);
\end{verbatim}
\end{quote}

The main steps of the program are:

\begin{enumerate}
    \item Create a connection with the server, with the {\it Host} and {\it Port} properties, and open the connection.
    \item Create a {\it socket stream} (an object representing the stream of data to be sent or received) and write the HEAD command; to write in a stream is equivalent to send that string of information.
    \item Read the answer of the server in the stream object.
\end{enumerate}

\subsection{Get the Head of a resource}

\subsubsection{C++ Example}

The following code is for downloading the Head of a resource from a server, using the {\it TNMHTTP} component (called here {\it MyHtml}) to make the source code more compact:

\begin{quote}
\begin{verbatim}
   try {
      MyHtml->Head( MyURL );
   }
   catch (Exception &E) {
      MyAnswer = E.Message;
      return;
   }
   MyAnswer = MyHtml->Header;
\end{verbatim}
\end{quote}

Note that the URL address of the resource is passed via the {\it MyURL} string, while the server response if saved in {\it MyAnswer}. Moreover, the {\it try-catch} statements are included to manage any possible exception that may arise from the connection operation.

\subsubsection{Perl Example}

The same problem as the previous one (downloading the Head information from a server) is resolved here with a Perl code and the {\it libwww-perl} package:

\begin{quote}
\begin{verbatim}
#!/usr/bin/perl -w
use strict;
use LWP::Simple;

my $url = 'www.someurl.com/index.html';
my $content = head($url);
die "Error" unless defined $content;
\end{verbatim}
\end{quote}

\subsection{Get a document in Java}

The following example code gets the content of an URL direction and write it on the screen:

\begin{quote}
\begin{verbatim}
import java.net.*;
import java.io.*;

class MyJava {
  public static void main( String[] args ) {
    String MyString;

    try {
      URL url = new URL(
         "http://www.mypage.com" );
          
      BufferedReader pageHtml = 
        new BufferedReader( new 
           InputStreamReader( url.openStream() ) );

      while( (MyString = pageHtml.readLine()) != null ) {
        System.out.println( MyString );
      }
    } catch( UnknownHostException e ) {
      e.printStackTrace();
      System.out.println( 
        "I/O Error" );
    } catch( MalformedURLException e ) {
      e.printStackTrace();
    } catch( IOException e ) { 
      e.printStackTrace();
    }
  }
}
\end{verbatim}
\end{quote}

\subsubsection{Get an image in Java}

Get an image in Java is very simple, thanks to the {\it Java Advanced Imaging (JAI) API}. The following code would load in the {\it image} structure the picture at {\verbºwww.mypage.com/myimage.gifº}:

\begin{quote}
\begin{verbatim}
import java.net.*;
import java.io.*;
import com.sun.media.jai.codec.*;
import com.sun.media.jai.codecimpl.*;

    URL url = "http://www.mypage.com/myimage.gif";
    RenderedImage image = JAI.create("url", url);
\end{verbatim}
\end{quote}

\subsection{Client identity}

In this example, we want to download the resource specified, but we want the server to know who we are: that is, we include a contact e-mail and the name of the agent used.

\subsubsection{C++ Example}

\begin{quote}
\begin{verbatim}
THeaderInfo *MyInfo;

   MyInfo = new THeaderInfo();
   MyInfo->Cookie = "";
   MyInfo->LocalMailAddress = "noname@nowhere.com";
   MyInfo->LocalProgram = "Internet Explorer";

   try {
      MyHtml->Get( MyURL );
   }
   catch (Exception &E) {
      ...
   }
   ...
\end{verbatim}
\end{quote}

\subsubsection{Perl Example}

\begin{quote}
\begin{verbatim}
$response = $browser->get($url,
    'User-Agent' => 'Internet Explorer',
    'From' => 'noname@nowhere.com',
       );
\end{verbatim}
\end{quote}

\subsection{People age logger in C++}

This is a more complex example. We know a web site where people can register and put their personal information: we are interested in making some statistics about the age distribution of the users.

We have the following information; the server is located at \verbºwww.server.comº, and the users profiles are at page \verbºwww.server.com/users.php?ID=1º (where $ID$ is the user number, i.e. $ID=1$, $ID=2$, $ID=3$, etc\ldots). Moreover, the age of the user is in the HTML page always after the following structure: \verbº<b>Age</font>º. The output should be saved in a text file.

The source code would be the following:

\begin{quote}
\begin{verbatim}
   for(UCurr = UStart; UCurr < UEnd; UCurr ++){
      bool Success;
      for(;;){
         try {
            NMHTTP1->Get( AnsiString( 
               "www.server.com/users.php?ID=") + 
               AnsiString(UCurr) );

            Success = true;
         }
         catch (...) {
            Success = false;
         }

         if(Success) break;
         Sleep(2000);
      }

      AnsiString Content;
      Content = NMHTTP1->Body;

      int Pos;
      Pos = Contenido.AnsiPos("<b>Age</font>");
      if(Pos != 0){
         for(i=0; i<4; i++) Age[i] = Content[Pos + i + 1];
         for(i=0; i<4; i++)
            if(Age[i] < '0' || Age[i] > '9') Age[i] = 0;
         Age[3] = 0;
      }
      if(strlen(Age) == 0) strcpy(Age, "0");

      out = fopen(OutputFile.c_str(), "at");
      fprintf(out, "%06d\t%s\n", UCurr, Age);
      fclose(out);

      Sleep(1000);
   }
\end{verbatim}
\end{quote}

Here there are some notes about how it works:

\begin{itemize}
    \item {\it UCurr} is the current user being studied; the loops runs from {\it UStart} to {\it UEnd}.
    \item For each user, there is a loop that try to download the page containing the age information; if some problems arises during the communication, the program waits 2 second and then try again.
    \item In the body retrieved, the program search for the \verbº<b>Age</font>º string: if it is found, the program extract the age supposing that it is codified in the next 3 characters of the string.
    \item Finally, the result (or $0$ if any) is saved in the {\it Output} file.
\end{itemize}

\subsection{Music lists in Perl}

Now, we will see a more complex example in Perl. This time, we have a web page (at \verbºwww.musicpage.comº) where you can search for a music artist, and you get other artists names and their {\it similarity} with the first. The search can be done at \verbºwww.musicpage.com/seach.php?a1=Nameº, where Name is the artist's name (you can specify up to 3 group names). The source code of the program will be the following:

\begin{quote}
\begin{verbatim}
#!/usr/bin/perl -w
use strict; $|++;

eval("use LWP 5.6.9;"); 
   die "Error: LWP required!" if $@;

my $base_url = "http://www.musicpage.com".
                  "/search.php";
my $counter = 0;
my $max_counter = 10;
my ($a1, $a2, $a3) = '';

$a1 = $ARGV[0] || die "No artists passed!\n";
$a2 = $ARGV[1] || "";
$a3 = $ARGV[2] || "";

print "Retrieving data...\n";
my $ua = LWP::UserAgent->new(agent => 'Internet Explorer');
my $data = $ua->get( 
   "$base_url&a1=$a1&a2=$a2&a3=$a3" )->content;

while ( $counter < $max_count &&
   $data =~ /href="art_info&artist=[^"]+">([^<]+)<\/a>[^<]+
          <\/td><td[^>]+><img[^>]+\/><img[^>]+
          width="([0-9]+)">(.*)/ ){

   print "%1.2f", ($2 / 300);
   print "\t\t" . $1 . "\n";
   $data = $3; $counter ++;
}

if ( $counter == 0 ) {print "No matches.\n";}
\end{verbatim}
\end{quote}

\section{FTP Protocol}

The FTP standard, like the HTTP, is an application protocol that handles a specific application: it uploads and downloads files from a server. There are always two computers involved in a FTP communication:

\begin{itemize}
    \item A {\it server}, that host the target files; it listens to the network for any incoming communication, it accepts connections (or reject them, according to some authentication rules) and manages the requests coming from the {\it clients}.
    \item A {\it client}, that initiates a connection to the server. Once connected, the client can request operations related with file manipulation: uploading files to the server, download files, rename or delete files and so on.
\end{itemize}

When programming a spider, the normal operation would involve HTTP communications; nevertheless, from time to time it may be necessary to download a file from a FTP server, so it make sense to see some communication examples.

In order to avoid a complete protocol analysis and implementation study, we are going to see examples with an FTP library, which automatically manages the connection; specifically, it will be the standard {\it TNMFTP} of Borland{\texttrademark} for the {\it C++} language.

\subsubsection{Download a file from a server}

The following is a simple example of a file download from a server:

\begin{quote}
\begin{verbatim}
   NMFTP1->Host = "www.thehost.com";
   NMFTP1->UserID = "MyName";
   NMFTP1->Password = "MyPassword";
   NMFTP1->Connect();

   NMFTP1->Mode(MODE_ASCII);
   NMFTP1->Download("TheRemoteFile.txt", "TheLocalFile.txt");
\end{verbatim}
\end{quote}

Let's see what it does:

\begin{itemize}
    \item {\it Host} is the URL of the server;
    \item {\it UserID} and {\it Password} is used to specify the authentication options, respectively the name of the user and his password; some servers admit a login without user name: in that case, {\it UserID} should be {\it anonymous}.
    \item {\it Connect()} open the connection with the server; in this example, we are supposing that the server is on-line and that a connection is always possible.
    \item {\it Mode}: there are several ways (or {\it modes}) to send (and receive) a file, depending on the content of the same; the most important are {\it ASCII} (to transmit text files) and {\it Binary} for programs and other files. Other two modes are defined in the standard, but are rarely used: {\it EBCDIC} mode and {\it Local} mode.
    \item {\it Download()}: this command tries to download the {\it remote file} to the local machine, saving it with the name {\it local file}.
\end{itemize}

In a real application, we cannot be sure of the status of the server: if it is down, or if it doesn't recognize the authentication, an error would be generated and the transaction would be aborted. To better handle this kind of situation, the previous code should be expanded with an {\it exception handling} code:

\begin{quote}
\begin{verbatim}
   NMFTP1->Host = "www.thehost.com";
   NMFTP1->UserID = "MyName";
   NMFTP1->Password = "MyPassword";
   try {
      NMFTP1->Connect();
   }
   catch (Exception &E) {
      Application->MessageBox(E.Message, "I/O Error", MB_OK);
   }
\end{verbatim}
\end{quote}

Depending on the result of the operation, an event would be generated: if this command succeeds, the {\it OnSuccess} event will be called, otherwise the {\it OnFailure} event is called.

\section{Background reading for Robots developers}

If you want to know more about robots developing and managing, there are some interesting books and pages in the WWW.

First, there are some books:

\begin{itemize}
    \item {\it Bots and Other Internet Beasties} by Joseph Williams, published by Pearson Education, 1996. ISBN 1575210169. 

    \item {\it Client Programming with Perl} by Clinton Wong. This book is now out of print, but is freely available through the O'Reilly Open Books Project (at \\ \verbºwww.oreilly.com/openbook/webclient/º).
 
    \item {\it Internet Agents: Spiders, Wanderers, Brokers, and Bots} by Fah-Chun Cheong, published by New Riders, 1995. ISBN 1-56205-463-5. 

    \item {\it Perl \& LWP} by Sean M. Burke, published by O'Reilly, 2002. ISBN 0596001789.

    \item {\it Spidering Hacks} by Kevin Hemenway, Tara Calishain, published by O'Reilly.
\end{itemize}

\vspace{0.5cm}

The {\it robotstxt} site has a great collection of papers related with spiders:

\begin{itemize}
    \item {\it Robots in the Web: threat or treat?} by Martijn Koster, at \\ \verbºwww.robotstxt.org/wc/threat-or-treat.htmlº.

    \item {\it Guidelines for Robot Writers} by Martijn Koster, at \\ \verbºwww.robotstxt.org/wc/guidelines.htmlº.

    \item {\it Evaluation of the Standard for Robots Exclusion} by Martijn Koster, at \\ \verbºwww.robotstxt.org/wc/eval.htmlº.
\end{itemize}

\vspace{0.5cm}

Other web pages that can be useful:

\begin{itemize}
    \item {\it Web Admin's Guide to Site Search Tools} at \\ \verbºwww.searchtools.com/guide/index.htmlº.

    \item {\it Web-based data mining} at \\ \verbºwww.ibm.com/developerworks/library/wa-wbdm/º.

    \item {\it Writing a Web Crawler in the Java Programming Language} at \\ \verbºjava.sun.com/developer/technicalArticles/ThirdParty/WebCrawler/º.

    \item {\it The RBSE Spider - Balancing Effective Search Against Web Load} at \\ \verbºmingo.info-science.uiowa.edu/eichmann/www94/Spider.psº.
\end{itemize}

\newpage

\section{Appendix}

\subsection{HTTP Response codes}

\begin{table}[!ht]
\centering
\begin{tabular}{|p{0.5in}|p{1.4in}|p{1.8in}|}
\hline\hline
{\emph Digit} & {\emph Type} & {\emph Description} \\
\hline
1xx & Informational & Only allowed in HTTP 1.1\\
2xx & Successful & The request was successfully received and processed \\
3xx & Redirection & Further actions are needed to complete the request \\
4xx & Client Error & The request contains bad syntax or cannot be fulfilled \\
5xx & Server Error & The server failed processing the request \\
\hline
\end{tabular}
\caption{HTTP Response code classes}
\label{tab:TabRC}
\end{table}

\begin{table}[!ht]
\centering
\begin{tabular}{|p{0.5in}|p{1.4in}|p{1.8in}|}
\hline\hline
{\emph Digit} & {\emph Code} & {\emph Description} \\
\hline
100 & Continue & The initial part of the request has been received, and the client may continue with its request. \\
101 & Switching Protocols & The server is complying with a client request to switch protocols to the one specified in the Upgrade header field. \\
\hline
\end{tabular}
\caption{1xx response codes}
\label{tab:TabRC1xx}
\end{table}

\begin{table}[!ht]
\centering
\begin{tabular}{|p{0.5in}|p{1.4in}|p{1.8in}|}
\hline\hline
{\emph Digit} & {\emph Code} & {\emph Description} \\
\hline
200 & OK & The client's request was successful, and the server's response contains the requested data. \\
201 & Created & This status code is used whenever a new URL is created. With this result code, the Location header (see next sections) is given by the server to specify where the new data was placed. \\
202 & Accepted & The request was accepted but not immediately acted upon. More information about the transaction may be given in the entity-body of the server's response. There is no guarantee that the server will actually honor the request, even though it may seem like a legitimate request at the time of acceptance. \\
203 & Non-Authoritative Information & The information in the entity header is from a local or third-party copy, not from the original server. \\
204 & No Content & A status code and header are given in the response, but there is no entity-body in the reply. Browsers should not update their document view upon receiving this response. This is a useful code for CGI programs to use when they accept data from a form but want the browser view to stay at the form. \\
\\
\hline
\end{tabular}
\caption{2xx response codes (first part)}
\label{tab:TabRC2xx}
\end{table}

\begin{table}[!ht]
\centering
\begin{tabular}{|p{0.5in}|p{1.4in}|p{1.8in}|}
\hline\hline
{\emph Digit} & {\emph Code} & {\emph Description} \\
\hline
205 & Reset Content & The browser should clear the form used for this transaction for additional input. Appropriate for data-entry CGI applications.\\
206 & Partial Content & The server is returning partial data of the size requested. Used in response to a request specifying a Range header. The server must specify the range included in the response with the Content-Range header. \\
\hline
\end{tabular}
\caption{2xx response codes (second part)}
\label{tab:TabRC2xx2}
\end{table}

\begin{table}[!ht]
\centering
\begin{tabular}{|p{0.5in}|p{1.4in}|p{1.8in}|}
\hline\hline
{\emph Digit} & {\emph Code} & {\emph Description} \\
\hline
300 & Multiple Choices & The requested URL refers to more than one resource. For example, the URL could refer to a document that has been translated into many languages. The entity-body returned by the server could have a list of more specific data about how to choose the correct resource. The client should allow the user to select from the list of URLs returned by the server, where appropriate. \\
301 & Moved Permanently & The requested URL is no longer used by the server, and the operation specified in the request was not performed. The new location for the requested document is specified in the Location header. All future requests for the document should use the new URL. \\
\hline
\end{tabular}
\caption{3xx response codes (first part)}
\label{tab:TabRC3xx}
\end{table}

\begin{table}[!ht]
\centering
\begin{tabular}{|p{0.5in}|p{1.4in}|p{1.8in}|}
\hline\hline
{\emph Digit} & {\emph Code} & {\emph Description} \\
\hline
302 & Moved Temporarily & The requested URL has moved, but only temporarily. The Location header points to the new location. Immediately after receiving this status code, the client should use the new URL to resolve the request, but the old URL should be used for all future requests. \\
303 & See Other & The requested URL can be found at a different URL (specified in the Location header) and should be retrieved by a GET on that resource. \\
304 & Not Modified & This is the response code to an If-Modified-Since header, where the URL has not been modified since the specified date. The entity-body is not sent, and the client should use its own local copy (see next sections). \\
305 & Use Proxy & The requested URL must be accessed through the proxy in the Location header. \\
\hline
\end{tabular}
\caption{3xx response codes (second part)}
\label{tab:TabRC3xx2}
\end{table}

\begin{table}[!ht]
\centering
\begin{tabular}{|p{0.5in}|p{1.4in}|p{1.8in}|}
\hline\hline
{\emph Digit} & {\emph Code} & {\emph Description} \\
\hline
400 & Bad Request & This response code indicates that the server detected a syntax error in the client's request. \\
401 & Unauthorized & The result code is given along with the WWW-Authenticate header to indicate that the request lacked proper authorization, and the client should supply proper authorization when requesting this URL again. \\
402 & Payment Required & This code is not yet implemented in HTTP. \\
403 & Forbidden & The request was denied for a reason the server does not want to (or has no means to) indicate to the client. \\
404 & Not Found & The document at the specified URL does not exist. \\
405 & Method Not Allowed & This code is given with the Allow header and indicates that the method used by the client is not supported for this URL. \\
406 & Not Acceptable & The URL specified by the client exists, but not in a format preferred by the client. Along with this code, the server provides the Content-Language, Content-Encoding, and Content-type headers. \\
407 & Proxy Authentication Required & The proxy server needs to authorize the request before forwarding it. Used with the Proxy-Authenticate header. \\
408 & Request Time-out & This response code means the client did not produce a full request within some predetermined time (usually specified in the server's configuration), and the server is disconnecting the network connection. \\
\hline
\end{tabular}
\caption{4xx response codes (first part)}
\label{tab:TabRC4xx}
\end{table}

\begin{table}[!ht]
\centering
\begin{tabular}{|p{0.5in}|p{1.4in}|p{1.8in}|}
\hline\hline
{\emph Digit} & {\emph Code} & {\emph Description} \\
\hline
409 & Conflict & This code indicates that the request conflicts with another request or with the server's configuration. Information about the conflict should be returned in the data portion of the reply. For example, this response code could be given when a client's request would cause integrity problems in a database. \\
410 & Gone & This code indicates that the requested URL no longer exists and has been permanently removed from the server. \\
411 & Length Required & The server will not accept the request without a Content-Length header supplied in the request. \\
412 & Precondition Failed & The condition specified by one or more If... headers in the request evaluated to false. \\
413 & Request Entity Too Large & The server will not process the request because its entity-body is too large. \\
414 & Request Too Long & The server will not process the request because its request URL is too large. \\
415 & Unsupported Media Type & The server will not process the request because its entity-body is in an unsupported format. \\
\hline
\end{tabular}
\caption{4xx response codes (second part)}
\label{tab:TabRC4xx2}
\end{table}

\begin{table}[!ht]
\centering
\begin{tabular}{|p{0.5in}|p{1.4in}|p{1.8in}|}
\hline\hline
{\emph Digit} & {\emph Code} & {\emph Description} \\
\hline
500 & Internal Server Error & This code indicates that a part of the server (for example, a CGI program) has crashed or encountered a configuration error. \\
501 & Not Implemented & This code indicates that the client requested an action that cannot be performed by the server. \\
502 & Bad Gateway & This code indicates that the server (or proxy) encountered invalid responses from another server (or proxy). \\
503 & Service Unavailable & This code means that the service is temporarily unavailable, but should be restored in the future. If the server knows when it will be available again, a Retry-After header may also be supplied. \\
504 & Gateway Time-out & This response is like 408 (Request Time-out) except that a gateway or proxy has timed out. \\
505 & HTTP Version Not Supported & The server will not support the HTTP protocol version used in the request. \\
\hline
\end{tabular}
\caption{5xx response codes}
\label{tab:TabRC5xx}
\end{table}

\subsection{HTTP Headers}

The HTTP standard makes a distinction between four different types of headers: 

\begin{itemize}
    \item General headers indicate general information such as the date, or whether the connection should be maintained. Both clients and servers can use them.
    \item Request headers are used only for client requests. They convey the client's configuration and desired document format to the server.
    \item Response headers are used only in server responses. They describe the server's configuration and special information about the requested URL. 
    \item Entity headers describe the document format of the data being sent between client and server. Although {\it Entity} headers are most commonly used by the server when returning a requested document, they are also used by clients when using the POST or PUT methods.
\end{itemize}

\begin{table}[!ht]
\centering
\begin{tabular}{|p{1.8in}|p{2.8in}|}
\hline\hline
{\emph Label} & {\emph Description} \\
\hline
Cache-Control & Specifies behavior for caching \\
Connection & Indicates whether network connection should close after this connection \\
Date & Specifies the current date \\
MIME-Version & Specifies the version of MIME used in the HTTP transaction \\
Pragma & Specifies directives to a proxy system \\
Transfer-Encoding & Indicates what type of transformation has been applied to the message body for safe transfer \\
Upgrade & Specifies the preferred communication protocols \\
Via & Used by gateways and proxies to indicate the protocols and hosts that processed the transaction between client and server \\
\hline
\end{tabular}
\caption{General headers}
\label{tab:GHeader}
\end{table}

\begin{table}[!ht]
\centering
\begin{tabular}{|p{1.8in}|p{2.8in}|}
\hline\hline
{\emph Label} & {\emph Description} \\
\hline
Accept & Specifies media formats that the client can accept \\
Accept-Charset & Tells the server the types of character sets that the client can handle \\
Accept-Encoding & Specifies the encoding schemes that the client can accept, such as compress or gzip \\
Accept-Language & Specifies the language in which the client prefers the data \\
Authorization & Used to request restricted documents \\
Cookie & Used to convey name=value pairs stored for the server \\
From & Indicates the email address of the user executing the client \\
Host & Specifies the host and port number that the client connected to. This header is required for all clients in HTTP 1.1. \\
If-Modified-Since & Requests the document only if newer than the specified date \\
If-Match & Requests the document only if it matches the given entity tags \\
If-None-Match & Requests the document only if it does not match the given entity tags \\
If-Range & Requests only the portion of the document that is missing, if it has not been changed \\
If-Unmodified-Since & Requests the document only if it has not been changed since the given date \\
Max-Forwards & Limits the number of proxies or gateways that can forward the request \\
Proxy-Authorization & Used to identify client to a proxy requiring authorization \\
Range & Specifies only the specified partial portion of the document \\
Referer & Specifies the URL of the document that contained the link to this one (i.e., the previous document) \\
User-Agent & Identifies the client program \\
\hline
\end{tabular}
\caption{Client (or Request) headers}
\label{tab:CHeader}
\end{table}

\begin{table}[!ht]
\centering
\begin{tabular}{|p{1.8in}|p{2.8in}|}
\hline\hline
{\emph Label} & {\emph Description} \\
\hline
Accept-Ranges & Declares whether or not the server accepts range requests, and if so, what units \\
Age & Indicates the age of the document in seconds \\
Proxy-Authenticate & Declares the authentication scheme and realm for the proxy \\
Public & Contains a comma-separated list of supported methods other than those specified in HTTP/1.0 \\
Retry-After & Specifies either the number of seconds or a date after which the server becomes available again \\
Server & Specifies the name and version number of the server \\
Set-Cookie & Defines a name=value pair to be associated with this URL \\
Vary & Specifies that the document may vary according to the value of the specified headers \\
Warning & Gives additional information about the response, for use by caching proxies \\
WWW-Authenticate & Specifies the authorization type and the realm of the authorization \\
\hline
\end{tabular}
\caption{Server (or Response) headers}
\label{tab:SHeader}
\end{table}

\begin{table}[!ht]
\centering
\begin{tabular}{|p{1.8in}|p{2.80in}|}
\hline\hline
{\emph Label} & {\emph Description} \\
\hline
Allow & Lists valid methods that can be used with a URL \\
Content-Base & Specifies the base URL for resolving relative URLs \\
Content-Encoding & Specifies the encoding scheme used for the entity \\
Content-Language & Specifies the language used in the document being returned \\
Content-Length & Specifies the length of the entity \\
Content-Location & Contains the URL for the entity, when a document might have several different locations \\
Content-MD5 & Contains a MD5 digest of the data \\
Content-Range & When a partial document is being sent in response to a Range header, specifies where the data should be inserted \\
Content-Transfer-Encoding & Identifies the transfer encoding used in the document \\
Content-Type & Specifies the media type of the entity \\
Etag & Gives an entity tag for the document \\
Expires & Gives a date and time that the contents may change \\
Last-Modified & Gives the date and time that the entity last changed \\
Location & Specifies the location of a created or moved document \\
URI & A more generalized version of the Location header \\
\hline
\end{tabular}
\caption{Entity headers}
\label{tab:EHeader}
\end{table}

\end{document}